%% file: main.tex
\documentclass[
  journal=pasa,
  manuscript=research-paper, %% or "review"
  year=202X,
  volume=YY,
]{cup-journal}

\usepackage{xcolor} 
\usepackage{hyperref}
\hypersetup{
    colorlinks=true,
    linkcolor=blue,
    citecolor=violet,
    urlcolor=teal
}
\usepackage{longtable}
\usepackage{amsmath}
\usepackage{cprotect}

\usepackage{graphicx}
\usepackage{overpic}
\usepackage{subcaption}
\usepackage[capitalise]{cleveref}
\usepackage{amssymb,microtype,siunitx,booktabs}
\crefname{figure}{Figure}{Figures}
\crefname{table}{Table}{Tables}
\crefname{equation}{Equation}{Equations}

\usepackage{multirow}
\usepackage[T1]{fontenc}
\usepackage{ae,aecompl}

\usepackage[separate-uncertainty=true,multi-part-units=single]{siunitx}
\usepackage{threeparttable}
\usepackage{outlines}
\newcommand{\citeg}[1]{\citep[e.g.][]{#1}}
\usepackage[normalem]{ulem}

\newcommand{\source}{ASKAP~J142431.2--612611}
\newcommand{\lpt}{ASKAP~J1424}

\newcommand{\preciseperiod}{\SI{2147.27}{\second}}

{%
  \end{outline}
}

\newcommand{\farcs}{\mbox{$^{\prime\prime}$}}

\DeclareSIUnit\jansky{Jy}
\DeclareSIUnit\mas{mas}
\DeclareSIUnit\year{yr}
\DeclareSIUnit\au{AU}
\DeclareSIUnit\parsec{pc}
\DeclareSIUnit{\dmunit}{\parsec\per\centi\meter\cubed}
\DeclareSIUnit\erg{erg}
\DeclareSIUnit\year{yr}
\DeclareSIUnit\gauss{G}
\DeclareSIUnit\percent{\%}
\DeclareSIUnit\beam{beam}
\DeclareSIUnit\angstrom{\text {Å}}
\DeclareSIUnit{\radlum}{\erg\per\second\per\hertz}
\DeclareSIUnit{\mjpb}{\milli\jansky\per\beam}
\DeclareSIUnit{\Rsun}{R_\odot}
\DeclareSIUnit{\Rjup}{R_{\rm J}}
\DeclareSIUnit{\Mjup}{M_{\rm J}}

\title{Discovery of a 36-minute long-period transient ASKAP J142431.2--612611}

\author{Joshua Pritchard}
\affiliation{CSIRO Space and Astronomy, PO Box 76, Epping, NSW 1710, Australia}

\author{Tara Murphy}
\affiliation{Sydney Institute for Astronomy, School of Physics, The University of Sydney, NSW 2006, Australia}
\alsoaffiliation{ARC Centre of Excellence for Gravitational Wave Discovery (OzGrav), Australia}

\author{Dougal Dobie}
\affiliation{Sydney Institute for Astronomy, School of Physics, The University of Sydney, NSW 2006, Australia}
\alsoaffiliation{ARC Centre of Excellence for Gravitational Wave Discovery (OzGrav), Australia}

\author{Emil Lenc}
\affiliation{CSIRO Space and Astronomy, PO Box 76, Epping, NSW 1710, Australia}

\author{Akash Anumarlapudi}
\affiliation{Department of Physics and Astronomy, University of North Carolina at Chapel Hill, 120 E. Cameron Ave, Chapel Hill, NC, 27599, USA}
\alsoaffiliation{Department of Physics, University of Wisconsin-Milwaukee, P.O. Box 413, Milwaukee, WI 53201, USA}

\author{Manisha Caleb}
\affiliation{Sydney Institute for Astronomy, School of Physics, The University of Sydney, NSW 2006, Australia}
\alsoaffiliation{ARC Centre of Excellence for Gravitational Wave Discovery (OzGrav), Australia}

\author{Sophia Grainger}
\affiliation{School of Mathematical and Physical Sciences, 12 Wally's Walk, Macquarie University, NSW 2109, Australia}
\alsoaffiliation{Macquarie University Astrophysics and Space Technologies Research Centre, Sydney, NSW 2109, Australia}
\alsoaffiliation{CSIRO Space and Astronomy, PO Box 76, Epping, NSW 1710, Australia}

\author{Natasha Hurley-Walker}
\affiliation{International Centre for Radio Astronomy Research, Curtin University, Bentley, WA, Australia}

\author{David L. Kaplan}
\affiliation{Department of Physics, University of Wisconsin-Milwaukee, P.O. Box 413, Milwaukee, WI 53201, USA}

\author{Samuel J. McSweeney}
\affiliation{International Centre for Radio Astronomy Research, Curtin University, Bentley, WA, Australia}

\author{Jackson Mitchell-Bolton}
\affiliation{Sydney Institute for Astronomy, School of Physics, The University of Sydney, NSW 2006, Australia}
\alsoaffiliation{CSIRO Space and Astronomy, PO Box 76, Epping, NSW 1710, Australia}

\author{Kovi Rose}
\affiliation{Sydney Institute for Astronomy, School of Physics, The University of Sydney, NSW 2006, Australia}
\alsoaffiliation{CSIRO Space and Astronomy, PO Box 76, Epping, NSW 1710, Australia}

\author{Rahul Sengar}
\affiliation{Max Planck Institute for Gravitational Physics (Albert Einstein Institute), D-30167 Hannover, Germany}
\alsoaffiliation{Leibniz Universität Hannover, D-30167 Hannover, Germany}
\alsoaffiliation{Department of Physics, University of Wisconsin-Milwaukee, P.O. Box 413, Milwaukee, WI 53201, USA}

\author{Ziteng Wang}
\affiliation{International Centre for Radio Astronomy Research, Curtin University, Bentley, WA, Australia}

\author{Jayde Willingham}
\affiliation{School of Mathematical and Physical Sciences, 12 Wally's Walk, Macquarie University, NSW 2109, Australia}
\alsoaffiliation{Macquarie University Astrophysics and Space Technologies Research Centre, Sydney, NSW 2109, Australia}
\alsoaffiliation{CSIRO Space and Astronomy, PO Box 76, Epping, NSW 1710, Australia}
\author{Andrew Zic}
\affiliation{CSIRO Space and Astronomy, PO Box 76, Epping, NSW 1710, Australia}

\received {dd Mmm YYYY}
\revised  {dd Mmm YYYY}
\accepted {dd Mmm YYYY}
\published{22 September 202X}

\keywords{XXXXXXXXXXXXX}

\begin{document}

\begin{abstract}
  We report the discovery of a new long-period radio transient,
  \source, with a \SI{36}{minute} period, identified in the Australian
  SKA Pathfinder Evolutionary Map of the Universe survey. We detected
  pulsed emission from \source\ over a period of eight days during
  follow-up observations with the Australia Telescope Compact Array,
  after which the source appears to have switched off. No optical or
  near-infrared counterpart is detected in archival surveys or in
  targeted Gemini South FLAMINGOS-2 observations. During its active
  state, the source exhibits a stable pulse profile with fractional
  polarisation consistent with \SI{100}{\percent}, evolving from
  elliptically to linearly polarised and tracing a well-defined
  great-circle trajectory on the Poincar{\'e} sphere. We show that
  this behaviour is consistent with fully linearly polarised intrinsic
  emission modified by propagation through a linearly polarised
  birefringent medium. This discovery expands the known population of
  long-period transients and highlights the intermittent nature of
  their activity. We discuss the implications for proposed models of
  long-period transients and outline future observations needed to
  constrain the origin of their intermittency and polarisation
  properties.
\end{abstract}

\input{content.tex}

\bibliographystyle{pasa-mnras}
\bibliography{bibfile}

\input{appendix.tex}

\end{document}

%% file: content.tex
\section{Introduction}

Long-period radio transients (LPTs) are a recently discovered class of
radio sources characterised by coherent pulsed emission with pulse
durations of seconds to minutes, repeating periodically on timescales
of tens of minutes to several hours \citep[for examples,
see][]{Hurley-Walker2022, Hurley-Walker2023, Caleb2024, Dobie2024,
   LiD2024, WangZ2025, Anumarlapudi2025, Bloot2025, deRuiter2025, 
  Dong2025, LeeY2025, McSweeney2025b}. These timescales of periodicity
are orders of magnitude longer than the millisecond to second
timescales of typical pulsars, placing LPTs within or below the
`death-valley' of the period-period derivative diagram
\citep{Rea2024}. In this regime, spin-down powered decay of the
magnetic field cannot support the pair-production required to power
pulsar emission. Hence LPTs can not be explained by conventional
pulsar emission models, raising fundamental questions about how
coherent radio emission can be sustained on such timescales.

The small number of LPTs that have been identified to date \citep[$\sim$14
at the time of publication, depending on exactly how the class is
defined;][]{Rea2026} show a rich diversity of radio
properties. Some have been routinely active over decades
\citeg{Hurley-Walker2023} while others appear intermittent and only
remain detectable for short windows before switching off
\citeg{Hurley-Walker2022, Caleb2024, Dobie2024, McSweeney2025b}. The emission
often shows high degrees of circular and linear polarisation; yet
while some sources show evolution in the polarisation position angle
(PA) reminiscent of pulsars \citep{Dobie2024}, others display flat or
slowly varying PA profiles \citep{Caleb2024}. In some objects the
pulse morphology and polarisation remain stable over an active window
\citep{LeeY2025}, while in others these properties change from pulse
to pulse, showing a variety of behaviours such as emission mode
changes \citeg{Caleb2024}, narrowband structure
\citeg{Anumarlapudi2025}, timing glitches \citeg{Dong2025}, frequency
dependent polarisation conversion, and orthogonal mode jumps
\citeg{Men2025}.

Multi-wavelength detections are beginning to provide insight into the
astrophysical classification of LPTs, and as a result at least two
subclasses are emerging. Some LPTs have infrared, optical, or
ultraviolet counterparts suggestive of a white dwarf (WD)
main-sequence binary system \citep[e.g.][]{deRuiter2025,
  Hurley-Walker2024, Rodriguez2025b, Anumarlapudi2025,
  Bloot2025}. Others show behaviour more consistent with a neutron
star progenitor: such as ASKAP~J1832$-$0911 which produces
periodically pulsed X-ray emission \citep{WangZ2025}. However, the
diverse radio phenomenology observed so far does not align neatly with
these associations, and more research is required to understand and
explain the phenomena.

The discovery of LPTs has been enabled by widefield facilities, namely
the Australian SKA Pathfinder \citep[ASKAP;][]{Johnston2008,
  Hotan2021}, the Murchison Widefield Array
\citep[MWA;][]{Bowman2013}, the LOw-Frequency ARray
\citep[LOFAR;][]{vanHaarlem2013}, and the Canadian Hydrogen Intensity
Mapping Experiment \citep[CHIME;][]{CHIME2018}. The large
instantaneous fields of view of these instruments allows widefield
radio surveys to prioritise long dwell times or repeat sampling of the
sky with high cadence, probing regions of transient parameter space
that have been previously inaccessible. Discoveries have also been
aided by the development of novel search techniques such as
fast-imaging \citep[e.g.][]{Caleb2024} and circular polarisation
searches \citep[e.g.][]{Anumarlapudi2025, LeeY2025}, allowing these
new regions of parameter space to be explored.

In this paper we report the discovery of a new long-period transient,
\source\ (hereafter \lpt), with a \SI{36}{\minute} period, identified
via a circular polarisation search. The source exhibited stable pulsed
emission for eight days before becoming undetectable, limiting the
available data but highlighting its intermittent nature. We present
these observations to enable and encourage future monitoring and
follow-up of this object. In \cref{sec:obs} we describe the
observations and data analysis, and in \cref{sec:discussion} we
discuss the radio properties of the source in the context of the
growing population of LPTs.

\section{Observations and Data Analysis}\label{sec:obs}

\subsection{Detection and radio follow-up}\label{sec:followup}

We discovered \lpt\ in a circular polarisation search of a
\SI{10}{\hour} ASKAP observation, conducted on 2025-01-09 (scheduling
block SB70271) as part of the Evolutionary Map of the Universe
\citep[EMU;][]{Norris2011, Norris2021} survey. We identified \lpt\ in
the full-integration images with a flux density of \SI{1.50\pm
  0.01}{\mjpb} in Stokes~$I$ (total intensity) and \SI{-0.12\pm
  0.01}{\mjpb} in Stokes~$V$\footnote{Throughout this paper we adopt
  the IAU/IEEE definition of Stokes~$V$ \citep{Robishaw2018}.}, giving
a fractional circular polarisation of \SI{8}{\percent}.

Following the ASKAP discovery, we conducted a multi-facility radio
follow-up campaign with the Australia Telescope Compact Array
\citep[ATCA;][]{Wilson2011}, MeerKAT \citep{Jonas2009}, the Parkes
\SI{64}{\meter} radio telescope (\textit{Murriyang}), and the
Murchison Widefield Array \citep[MWA;][]{Tingay2013, Wayth2018}.  We
detected \lpt\ in both ATCA followup observations at
\SI{2100}{\mega\hertz}, but made no detections with any other
facility. The details of all radio followup observations are described
in \ref{apx:followup} and summarised in \cref{tab:observations}. We
also searched archival radio and multi-wavelength observations,
including 60 observations as part of the Rapid ASKAP Continuum Survey
\citep[RACS;][]{McConnell2020, Duchesne2023a} and Variables And Slow
Transients \citep[VAST;][]{Murphy2021} surveys with ASKAP, but made no
detections. Details and limits are provided in \ref{apx:archival}.

\begin{table}
  \centering
  \caption[Summary of radio followup observations]{
    \small
    Summary of radio observations of \lpt\ with ASKAP, ATCA, MWA, MeerKAT
    and \textit{Murriyang}. Columns are the observation start time,
    telescope, project code / SBID, frequency range $\nu_{\rm obs}$, and
    duration $t_{\rm obs}$.
  }\label{tab:observations}
    \begin{tabular}[t]{lcccc}
      \hline
      UT Start Time      & Telescope & Project Code & $\nu_{\rm obs}$ & $t_{\rm obs}$ \\
                         &         &               & [\si{\mega\hertz}] & [\si{\hour}] \\
      \hline
      2025--01--09 20:00 & ASKAP   & SB70271          & 800--1088  & 10   \\
      2025--01--14 18:30 & \textit{Murriyang} & P1050 & 704--4032  & 1    \\
      2025--01--15 18:30 & MWA     & G0115            & 185--215   & 3    \\
      2025--01--15 20:00 & ATCA    & C1726            & 1100--3100 & 1.5  \\
      2025--01--16 18:30 & MWA     & G0115            & 185--215   & 8    \\  
      2025--01--17 21:08 & ATCA    & C3363            & 1100--3100 & 6.1  \\
      2025--01--28 18:30 & \textit{Murriyang} & P1050 & 704--4032  & 2    \\
      2025--02--21 03:10 & MeerKAT & 20250224-0006    & 544--1088  & 4    \\
      2025--02--26 02:11 & MeerKAT & 20250217-0010    & 544--1088  & 4    \\
      2025--09--28 01:16 & ASKAP   & SB77270          & 800--1088  & 10   \\
      \hline
    \end{tabular}
\end{table} 

\subsection{Infrared photometric follow-up}

\begin{figure}
  \centering
  \includegraphics[width=\textwidth]{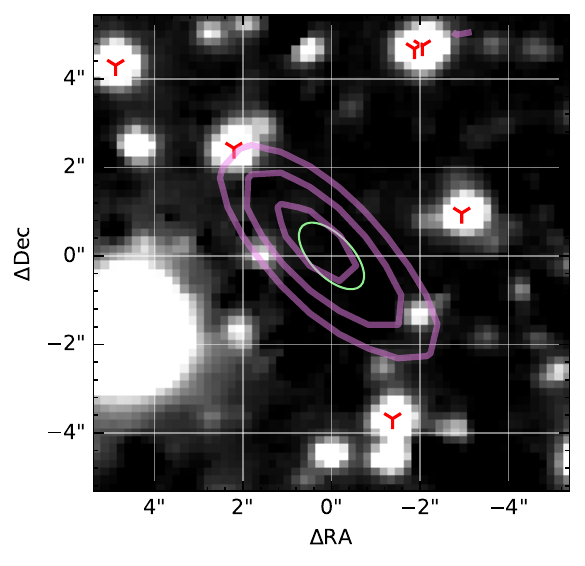}
  \caption[Gemini photometry]{
    \small
    $K_s$-band Gemini observation with Stokes~$I$ radio contours from
    ATCA C3363 observation overlaid. The green ellipse indicates the
    5$\sigma$ astrometric uncertainty of $0\farcs9-1\farcs9$. The 
    positions of catalogued VVV sources are indicated with red markers.
  }\label{fig:gemini}
\end{figure}

We acquired near infrared observations in the $J$ and $K_S$ bands with
the FLAMINGOS-2 instrument on Gemini South on 2025-01-21. We performed
standard data reduction using the \textsc{DRAGONS} pipeline
\citep{Labrie2023}, and astrometric calibration by extracting source
positions with \textsc{photutils} and registering against the
proper-motion corrected positions of stars in Gaia DR3
\citep{Gaia2023}. Photometric calibration was obtained by matching
isolated and unsaturated field stars to the Variables in the Via
Lactea survey \citep[VVV;][]{Minniti2010}, yielding zero-point
uncertainties of \SI{0.1}{mag} and \SI{0.15}{mag} in the $J$ and $K_S$
bands, respectively. In \cref{fig:gemini} we show a cutout image of
the $K_S$-band data with contours of our ATCA C3363 observation
overlaid. We did not detect a source within the $5\sigma$ astrometric
uncertainty of $0\farcs9-1\farcs9$ in either band, with 3$\sigma$
limiting magnitudes\footnote{All magnitudes are in the AB system.}  of
$J > 22.5$ and $K_S > 19.5$.

\subsection{Time-domain analysis}

We processed all calibrated radio observations by forming a model of
background sources of emission with \textsc{WSclean}
\citep{Offringa2014}, and extracting model-subtracted dynamic spectra
in all Stokes parameters using
\textsc{DStools}\footnote{\url{https://github.com/askap-vast/dstools}}
\citep{Pritchard2025a}. We then frequency-averaged the dynamic spectra
to form visibility lightcurves.

We fit Gaussian profiles to each detected pulse and used the
barycentred times of arrival to determine an ephemeris of the form
$\phi(t) = \frac{t - t_0}{P} + \phi_0$, with a best fit period
$P = \SI{2147.27 \pm 0.01}{\second}$. The reference time
$t_0 = {}$2025-01-09 20:32:11.701 corresponds to the barycentric time
of arrival of the first fully sampled pulse in SB70271, with a
reference pulse phase of $\phi_0 = 0.5$. The sequence of 17
consecutive pulses detected in this observation uniquely constrains
the period and rules out alias periods. Owing to the small time range
spanned by the 2025 detections, we are unable to meaningfully
constrain the source's spin-up or spin-down, and place an upper limit
of order $\SI{e-8}{\second\per\second}$ on the period derivative.

We used this ephemeris to phase-fold all radio observations.  Several
archival observations obtained between 2021-02-01 and 2024-10-03
sample a significant fraction of the estimated pulse window yet show
no emission, suggesting that \lpt\ was not active for multiple years
prior to the 2025-01-09 detection. However, it is possible that \lpt\
has short windows of activity as observed in several other LPTs
\citeg{Hurley-Walker2022, Caleb2024, Dobie2024, McSweeney2025b}, and
that such intervals were not sampled by the available archival
data. The phase-folded lightcurves for all $\sim\si{\giga\hertz}$
observations are presented in \ref{apx:followup}.

Phase-folded dynamic spectra of the 17 pulses detected in SB70271 show
frequency-dependent phase shifts across the folded pulse
profile. Interpreting these shifts as purely dispersive implies a
dispersion measure of $\mathrm{DM} = \SI{1400 \pm 100}{\dmunit}$.
However, intrinsic frequency-dependent pulse morphology can produce
similar phase offsets. When we allow the pulse profile to vary with
frequency while simultaneously fitting for dispersion, the inferred DM
decreases to $\SI{200 \pm 40}{\dmunit}$. As the phase shifts in these
data are small compared to the pulse width, dispersion and profile
evolution cannot be distinguished.  We therefore treat
$\SI{1400}{\dmunit}$ as a conservative upper limit on the DM under the
assumption that the observed phase shifts are entirely
dispersive. Details of the dispersion modelling are provided in
\ref{apx:dispersion}.

We measured the spectral index $\alpha_\nu$ of \lpt\ in these
observations by fitting a power law $S(\nu) \propto \nu^{\alpha_\nu}$
to the peak sample of the phase-folded lightcurves. All detections are
well-described by a power law, though $\alpha_\nu$ changes from $-1.6$ in
the \SIrange{800}{1088}{\mega\hertz} ASKAP band on 2025-01-09 to
$-1.2$ in the \SIrange{1100}{3100}{\mega\hertz} ATCA band on
2025-01-15. These observations also show a significant decrease in
band-averaged flux density from $\sim$\SI{65}{\milli\jansky} to
\SI{12}{\milli\jansky}, which cannot be explained solely by spectral
shape. As our MWA observations were acquired quasi-contemporaneously
with the ATCA observations, non-detection at \SI{200}{\mega\hertz}
suggests a limit on the \SIrange{200}{1100}{\mega\hertz} spectral index
$\alpha_\nu> -1$. It is therefore likely that \lpt\ has a spectral
turnover between \SIrange{200}{800}{\mega\hertz}, similar to
ASKAP~J1832$-09$ \citep{WangZ2025} and ASKAP~J1839$-07$
\citep{LeeY2025}.

\subsection{Polarisation properties}

\begin{figure}
  \centering
  \includegraphics[width=\textwidth]{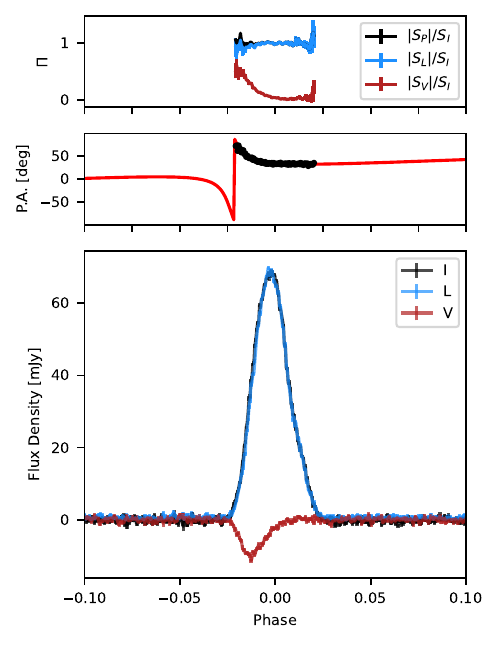}
  \caption[Folded pulse profile.]{
    \small
    Folded pulse profile of 17 pulses detected in ASKAP observation SB70271.
    Lightcurves are formed from frequency-averaged dynamic spectra
    binned to $2000$ pulse phase bins, folded at a period of \preciseperiod.
    From bottom to top, panels show full polarisation folded pulse profiles,
    polarisation position angle, and fractional polarisation. Points in the
    top two panels are masked below a signal-to-noise ratio (SNR) of
    $S_I/\sigma_{S_I} < 5$.
  }\label{fig:askap-folded-lc}
\end{figure}

We measured the Faraday rotation measure (RM) of all detected pulses
using RM synthesis \citep{Brentjens2005} with the
\textsc{RM-lite}\footnote{\url{https://github.com/AlecThomson/rm-lite}}
implementation of \textsc{RM-tools} \citep{Purcell2020}. All pulses
show a simple Faraday dispersion function with a single unresolved
peak at an RM of \SI{-222}{\radian\per\square\meter} and no evidence
for RM variations with pulse phase.  In \cref{fig:askap-folded-lc} we
show the phase-folded, RM-corrected pulse profile of 17 pulses from
the ASKAP SB70271 observation. The fractional polarisation is
consistent with \SI{100}{\percent} across the full pulse profile,
transitioning from an elliptical state during the first half of the
pulse to fully linearly polarised.

The PA begins at $\SI{70}{\deg}$ and flattens out at $\SI{30}{\deg}$
as the circular polarisation fraction decreases. The evolution in PA
is reminiscent of the S-shaped swing in pulsar PA profiles and
commonly explained by the rotating vector model
\citep[RVM;][]{Radhakrishnan1969}, in which the PA corresponds to the
projection of the local magnetic field orientation at the emission
site onto the plane of the sky. We used the Markov chain Monte Carlo
sampler \textsc{emcee} \citep{Foreman-Mackey2013} to fit an RVM
profile to the phase folded ASKAP SB70271 lightcurve with a best fit
of $\alpha = \SI{170 \pm 4}{\deg}$, $\beta = \SI{-0.3 \pm 0.1}{\deg}$,
$\phi_0 = \SI{-8.37 \pm 0.04}{\deg}$, and
$\psi_0 = \SI{-72.6 \pm 0.3}{\deg}$. Here $\psi(\phi)$ is the PA as a
function of pulse phase $\phi$, $\alpha$ is the angle between rotation
and magnetic axes, $\beta$ is the angle between the magnetic axis and 
line of sight, and $\psi_0$ and $\phi_0$ are the PA and phase
around which the PA swing is centred.

While the RVM provides a good empirical description of the observed PA
evolution, it is not clear that the emission mechanism powering \lpt\
produces radiation polarised in the sense of the local magnetic field,
as is typically assumed for pulsar emission, nor that the fitted RVM
parameters map directly onto a rotational geometry. The fitted offset
$\phi_0$ between steepest PA gradient and the total intensity peak is
also atypical of most pulsars, for which $\phi_0$ is generally
positive and attributed to aberration and retardation effects
associated with finite emission heights \citep{Blaskiewicz1991}.

The RVM also only accounts for the linearly polarised component
of the emission and does not describe the presence or evolution of
circular polarisation. To characterise the full polarisation behaviour
we therefore analysed the evolution of the normalised Stokes
parameters $(Q/P, U/P, V/P)$ on the Poincar{\'e} sphere, where
$P = \sqrt{Q^2 + U^2 + V^2}$. Full details of the Poincar{\'e} sphere
analysis are provided in \ref{apx:poincare}.

\begin{figure}
  \centering
  \includegraphics[width=\textwidth]{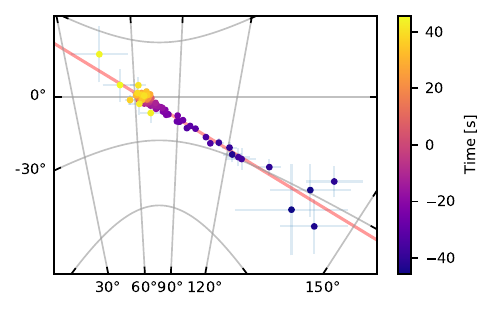}
  \caption[Poincare sphere projection.]{
    \small
    Time evolution of polarisation state in the ASKAP SB70271 folded
    pulse profile.  Points show a Gnomonic projection of Poincar{\'e}
    sphere normalised by total polarisation
    $P = \sqrt{Q^2 + U^2 + V^2}$. The red line shows the best linear fit
    to the projected data, corresponding to a great circle of
    inclination $\SI{31.5 \pm 0.6}{\deg}$ crossing the equator at
    $\SI{64.0 \pm 0.4}{\deg}$ longitude.
  }\label{fig:poincare-sphere}
\end{figure}

In \cref{fig:poincare-sphere} we show the polarisation state of the
folded ASKAP SB70271 pulse profile in a Gnomonic projection of the
Poincar{\'e} sphere, in which great circles are mapped to straight
lines. The data are well fit by a great circle with inclination angle
of $\delta = \SI{31.5 \pm 0.6}{\deg}$ and intersecting the equator of
the sphere at a longitude of
$2\psi_{\rm eq} = \SI{64.0 \pm 0.4}{\deg}$, corresponding to a PA of
$\psi_{\rm eq} = \SI{32.0 \pm 0.2}{\deg}$.  The great circle
trajectory shows only marginal evidence for weak frequency dependence,
with inclination angle varying with wavelength as
$\delta \propto \lambda^p$ with $p = 0.5 \pm 0.2$ when keeping
$2\psi_{\rm eq}$ fixed to $\SI{32.0}{\deg}$ and $p = 0.21 \pm 0.06$
when allowing it to vary freely. A full description of this model
is provided in \ref{apx:poincare-lbm}.

\section{Discussion and Conclusions}\label{sec:discussion}

We have presented the discovery of a long-period transient, \lpt, with
a \SI{36}{\minute} period that exhibits an extremely stable pulse
profile over an \SI{8}{\day} activity window, as observed with ASKAP
and ATCA across \SIrange{800}{3100}{\mega\hertz}. The emission is
consistent with being \SI{100}{\percent} polarised across the full
pulse profile, and evolves along a great circle trajectory on the
Poincar{\'e} sphere as a function of pulse phase transitioning from an
elliptical to fully linearly polarised state.

Evolution of the polarisation state along a great circle may be explained
by a fully linearly polarised intrinsic polarisation state which then
passes through a linear birefringent medium (LBM) in which the natural
modes are linearly polarised \citep{Lyutikov2022, Bera2025}. In the
Poincar{\'e} sphere picture, such a medium applies a fixed rotation to
the intrinsic polarisation state about an axis aligned with the LBM
natural modes, transforming an intrinsic equatorial great circle
trajectory into an inclined great circle. Evolution of the pulse
phase-resolved polarisation state along a great circle has been
previously observed from the LPT ASKAP~J1755-2527 \citep{Dobie2024},
though with lower fractional polarisation and possibly being driven
by partially coherent mixing of orthogonal modes \citep{Oswald2023a}.
Similar behaviour has also been observed in fast radio bursts
\citep{Bera2025}.

We have not identified an optical or infrared counterpart to \lpt\ in
either archival data or targeted follow-up observations. As is the
case for the majority of LPTs, \lpt\ is located at low Galactic
latitude ($b \sim \SI{-0.5}{\degree}$) where significant extinction is
expected, limiting the constraining power of our optical and infrared
limits on the nature of the progenitor. \citet{Horvath2026}
demonstrate that GPM~J1839$-$10 is likely a WD binary system, with
pulse activity driven by magnetic interaction when the WD magnetic
axis intersects the companion's magnetised wind, and recurring at the
beat between WD rotation and orbital periods. Considering similar
modulation of the activity of \lpt\, we constrain the activity window
of this system to be between \SIrange{8}{141}{\day} based on the ASKAP
and ATCA detections and nearest-in-time non-detections. Under the
\citet{Horvath2026} model this would imply a beat period of at least
several weeks, possibly months. The lack of detections across two
years of the VAST Galactic survey may be difficult to reconcile with
quasi-periodic activity windows on timescales of weeks. However this
is not a strong constraint, as each VAST observation only samples a
third of the period and the observing cadence is longer than the
nominal activity window.

Further monitoring (e.g. as part of the planned second phase of the
VAST Galactic survey) will allow us to determine whether the observed 
emission follows an intermittent activity pattern, or was powered by 
a one-off or stochastic event such as accretion of plasma from a 
companion. Continued characterisation of the polarisation dynamics of 
\lpt\ in these observations will also provide insight into the viability 
of an LBM model in explaining the great circle trajectory of the 
polarisation state, providing important insights into further
understanding the  emission mechanism and constraining the plasma 
properties of the near-source medium.

The second phase of the VAST Galactic survey will begin in early 2026
with a revised observing strategy consisting of several high-cadence
observing blocks per year. The VAST Galactic fields will be looped
over $\sim 10$ times in the space of a few days, supplementing the
fortnightly cadence executed in the first phase of the survey. This
strategy will provide stronger constraints on orbital modulation in
tight binary orbits, and will allow characterisation of pulse
evolution over timescales of days, such as the fading observed in
\lpt\ and mode switching observed in ASKAP~J1935 \citep{Caleb2024}.

\section*{Acknowledgements}

Parts of this research were supported by the Australian Research
Council Centre of Excellence for Gravitational Wave Discovery
(OzGrav), project number CE230100016.  DK was supported by NSF grant
AST-2511757.

% ASKAP
This scientific work uses data obtained from Inyarrimanha Ilgari
Bundara, the CSIRO Murchison Radio-astronomy Observatory. We
acknowledge the Wajarri Yamaji People as the Traditional Owners and
native title holders of the Observatory site. CSIRO’s ASKAP radio
telescope is part of the ATNF
\href{https://ror.org/05qajvd42}{Australia Telescope National
  Facility}. Operation of ASKAP is funded by the Australian Government
with support from the National Collaborative Research Infrastructure
Strategy. ASKAP uses the resources of the Pawsey Supercomputing
Research Centre. Establishment of ASKAP, Inyarrimanha Ilgari Bundara,
the CSIRO Murchison Radio-astronomy Observatory and the Pawsey
Supercomputing Research Centre are initiatives of the Australian
Government, with support from the Government of Western Australia and
the Science and Industry Endowment Fund.

% ATCA
The Australia Telescope Compact Array is part of the Australia
Telescope National Facility (ATNF,
\href{https://ror.org/05qajvd42}{Australia Telescope National
  Facility}) which is funded by the Australian Government for
operation as a National Facility managed by CSIRO. We acknowledge the
Gomeroi people as the Traditional Owners of the Observatory site.

% MWA
Support for the operation of the MWA is provided by the Australian
Government (NCRIS), under a contract to Curtin University administered
by Astronomy Australia Limited. ASVO has received funding from the
Australian Commonwealth Government through the National eResearch
Collaboration Tools and Resources (NeCTAR) Project, the Australian
National Data Service (ANDS), and the National Collaborative Research
Infrastructure Strategy.

% MeerKAT
The MeerKAT telescope is operated by the South African Radio Astronomy
Observatory, which is a facility of the National Research Foundation,
an agency of the Department of Science and Innovation.

% Gemini
Based on observations obtained at the international Gemini
Observatory, a program of NSF NOIRLab, which is managed by the
Association of Universities for Research in Astronomy (AURA) under a
cooperative agreement with the U.S. National Science Foundation on
behalf of the Gemini Observatory partnership: the U.S. National
Science Foundation (United States), National Research Council
(Canada), Agencia Nacional de Investigaci\'{o}n y Desarrollo (Chile),
Ministerio de Ciencia, Tecnolog\'{i}a e Innovaci\'{o}n (Argentina),
Minist\'{e}rio da Ci\^{e}ncia, Tecnologia, Inova\c{c}\~{o}es e
Comunica\c{c}\~{o}es (Brazil), and Korea Astronomy and Space Science
Institute (Republic of Korea).

\section*{Data Availability}
The ASKAP data used in this paper can be accessed through the CSIRO
ASKAP Science Data Archive
(CASDA\footnote{\url{https://data.csiro.au/domain/casdaObservation}})
under project codes AS106 and AS107.  The MWA data used in this paper
can be accessed through the All-Sky Virtual Observatory
(ASVO\footnote{\url{https://asvo.mwatelescope.org/}}) under project
codes G0080 and G0115.

%% file: appendix.tex
\appendix

\section{Radio followup observations}\label{apx:followup}
  
ATCA observations were acquired on 2025-01-17 (project code C1726) and
2025-01-22 (project code C3363) in the 6C configuration using the
\SI{16}{\centi\meter} (L-band) receiver with a \SI{2048}{\mega\hertz}
band centred on \SI{2100}{\mega\hertz}. We calibrated the bandpass
response, flux scale, and polarisation leakage with observations of
PKS~B1934$-$638 (PKS~J1939$-$6342), and solved for time-varying
antenna gains with interleaved \SI{90}{\second} scans of ATCA
calibrator 1352-63 (PMN~J1355$-$6326). We flagged and calibrated the
data using standard continuum data processing routines in
\textsc{miriad}. We detected \lpt\ in both observations with an
image-plane flux density of \SI{330 \pm 30}{\micro\jansky\per\beam}
and a best fit position of 14h24m31.40s $\pm 0\farcs4$ -61d26m10.8s
$\pm 0\farcs2$.

We obtained two MeerKAT observations for \SI{4}{\hour} each on
2025-02-21 and 2025-02-26 (project ID SCI-20241101-TM01) using the UHF
band receiver, with a \SI{544}{\mega\hertz} band centred on
\SI{816}{\mega\hertz}. We used the SARAO Science Data Processor (SDP)
pipeline to flag and calibrate the data, using PKS~J1939$-$6342 to
solve for the bandpass response, flux scale, and polarisation leakage,
J1446$-$4701 to solve for time-varying gains, and 3C~286 to calibrate
the absolute polarisation position angle.  We further corrected the
visibilities for mislabelling of the $X$ and $Y$ feeds
\citep{Perley2022} and parallactic angle rotation. We did not detect
\lpt\ in either observation, with 3$\sigma$ Stokes~$I$ sensitivity
limits of $\sim$\SI{100}{\micro\jansky\per\beam} in the
full-integration images and $\sim$\SI{1}{\milli\jansky} in
\SI{2}{\minute} cadence lightcurves.

We acquired \textit{Murriyang} observations 2025-01-14 and 2025-01-28
for durations of \SI{1}{\hour} and \SI{2}{\hour} respectively. We used
the ultra-wide-bandwidth low-frequency receiver
\citep[UWL;][]{Hobbs2020} which provides continuous frequency coverage
from \SI{704}{\mega\hertz} to \SI{4032}{\mega\hertz}, and recorded
data with a sampling rate of \SI{64}{\micro\second} with $6656$
\SI{500}{\kilo\hertz}-wide channels across the \SI{3328}{\mega\hertz}
bandwidth. We conducted standard pulsar searches on the data from both
epochs\footnote{Due to heavy winds, observations during the first
  epoch were taken in a non-contiguous manner.}. In addition, we also
conducted sub-banded searches, splitting the data into different
sub-bands and performing pulsar searches in each sub-band. No
pulsations were found in either epoch. We also examined the raw
voltage data around the predicted time of arrival of the pulse on
2025-01-14, but detected no significant excess corresponding to the
\SI{2}{\minute} pulse, even after cleaning the data for strong radio
frequency interference.

We performed MWA follow-up observations under project code G0115,
continuously tracking \lpt\ for \SI{3}{\hour} and \SI{8}{\hour} on the
nights of 2025-01-15 and 2025-01-16, respectively. We examined
\SI{5}{\minute} integration images, as well as \SI{4}{\second}
resolution dynamic spectra, towards the location of \lpt. No
detections were made, with 3$\sigma$ sensitivity limits of
$\sim$\SI{150}{\mjpb} in the \SI{5}{\minute} integrations, and
$\sim$\SI{1.5}{\jansky\per\beam} in the \SI{4}{\second} cadence light
curves.

\begin{figure*}[t!]
  \centering
  \includegraphics[width=\textwidth]{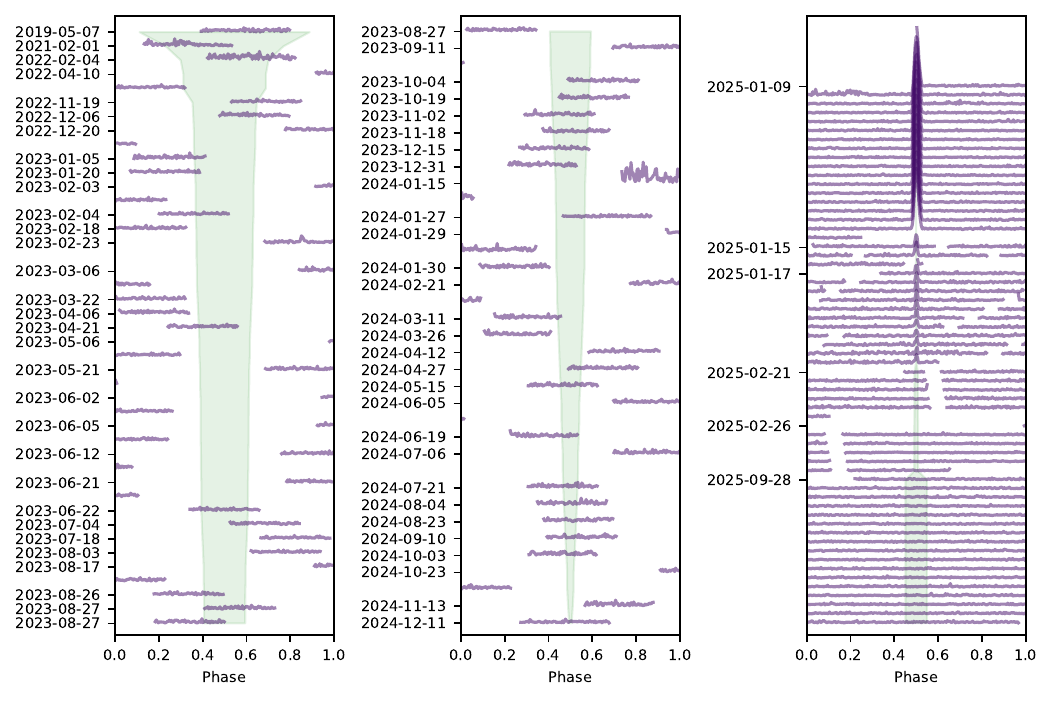}
  \caption[Phase folded archival observations.]{
    \small
    Lightcurves from all ASKAP, ATCA, and MeerKAT radio observations
    phase-folded to the radio period of \preciseperiod. Green
    shading indicates the range of uncertainty in predicted pulse time
    around the expected pulse phase of 0.5. Pulses are only detected
    in the ASKAP and ATCA observations between 2025-01-09 and 2025-01-17.
  }\label{fig:observations-stacked}
\end{figure*}

In \cref{fig:observations-stacked} we show the phase folded
lightcurves of all $\sim\si{\giga\hertz}$ radio observations.
 
\section{Archival multi-wavelength data}\label{apx:archival}

\lpt\ was covered by a second EMU observation (SB77270) on 2025-09-28
but no pulses were detected. Phase one of the VAST Galactic survey ran
for approximately two years from November 2022. Observations were
carried out on an approximately fortnightly basis, with each
observation lasting \SI{12}{\minute}. \lpt\ is within the survey
footprint and was observed 53 times across the survey. The position of
\lpt\ was also observed seven times across the multiple epochs of RACS
in the Low, Mid and High band.  \lpt\ was not detected in any VAST or
RACS archival observations.

We searched the Galactic Plane Monitoring programme undertaken with
the MWA, described in the Methods of \cite{Hurley-Walker2023};
585~observations sensitive to \lpt\ were taken under project code
G0080 over 2022 and 2024 Jun--Oct. These covered
\SIrange{185}{215}{\mega\hertz} over $|b|<\SI{15}{\deg}$ and
$\SI{285}{\deg} < l < \SI{65}{\deg}$, for \SIrange{30}{45}{\minute}
integration on a bi-weekly cadence (Hurley-Walker et al. in
preparation will describe the survey in full). No detections were
made, with 3$\sigma$ sensitivity limits of $\sim\SI{60}{\mjpb}$ in
the \SI{5}{\minute} integrations, and $\sim\SI{600}{\mjpb}$ in the
\SI{4}{\second} light curves.

\lpt\ has not been detected in any archival infrared (IR), optical,
ultraviolet (UV), or X-ray data. The strongest archival limits in each
band are $g>23.5$, $r>22.6$, $i>22.1$, $z>21.6$, and $Y>20.8$ from the
DECam Plane Survey \citep[DECAPS;][]{Saydjari2023}; $J>19.9$,
$H>18.1$, and $Ks>16.9$ from the VISTA Variables in the Via Lactea
\citep[VVV;][]{Minniti2010} survey; $UVM2 >21.67$ from a
\SI{0.46}{\kilo\second} observation with \textit{Neil Gehrels Swift
  Observatory} \citep[\textit{Swift;}][]{Gehrels2004} (ID
00040976009). We also find no detection in XMM Newton X-ray telescope
\citep{Jansen2001} observations in the \SIrange{0.2}{12}{\kilo\eV}
band above a 3$\sigma$ flux limit of
$F_X < \SI{1.9e-12}{\erg\,\second^{-1}\cm^{2}}$.

\section{Dispersion modelling}\label{apx:dispersion} 

We estimated the dispersion measure (DM) from the phase-folded SB70271
dynamic spectrum by measuring frequency-resolved times of arrival
across the \SIrange{744}{1032}{\mega\hertz} band. We fit a Gaussian of
varying amplitude and central phase to the folded profiles in
\SI{4}{\mega\hertz} subbands, keeping the width fixed to that of the
broadband pulse profile. We adopted the centre of the Gaussian model
as the time of arrival for each subband, and fit the
frequency-dependent phase offsets with the standard cold-plasma
dispersion law to obtain $\mathrm{DM} = \SI{1400 \pm 100}{\dmunit}$.

This approach assumes that the pulse shape does not vary with
frequency. When the dispersive delay across the band is comparable to
or smaller than the pulse width, intrinsic profile evolution becomes
degenerate with dispersion and can bias the inferred DM. To assess the
impact of profile evolution, we used \textsc{PulsePortraiture}
\citep{Pennucci2019} to construct a frequency-dependent pulse model
from the phase-folded SB70271 data, using three eigen-components to
describe the frequency-dependent morphology. We used \texttt{pat} from
\textsc{PSRCHIVE} \citep{Hotan2004} to measure times of arrival for
each channel of the profile across frequency using this
frequency-evolving model, and fit the DM using \textsc{TEMPO2}
\citep{Hobbs2006}. This joint modelling of profile variation and
dispersion results in a DM of $\SI{210 \pm 40}{\dmunit}$.
  
The substantial difference between these estimates highlights the
challenge of disentangling DM from profile frequency evolution.  Given
this degeneracy, we adopt $\SI{1400}{\dmunit}$ as a conservative upper
limit on the DM under the assumption the frequency-dependent phase
offsets are purely dispersive. A robust measurement will require
observations with higher time resolution capable of resolving pulse
sub-structure, together with broader simultaneous frequency coverage.

\section{Polarisation modelling on the Poincar{\'e} sphere}\label{apx:poincare}

\subsection{Poincar{\'e} sphere representation}

In the Poincar{\'e} sphere formalism the polarisation vector
\begin{equation}
  \mathbf{P} = \frac{1}{P}
  \begin{bmatrix}
    Q \\
    U \\
    V
  \end{bmatrix}
  =
  \begin{bmatrix}
    \cos{2\chi}\cos{2\psi} \\
    \cos{2\chi}\sin{2\psi} \\
    \sin{2\chi}
  \end{bmatrix}
\end{equation}
fully describes the polarisation state as a location on the surface of
a sphere, where the longitudinal coordinate $2\psi \in [-\pi, \pi]$
describes the PA and latitudinal coordinate
$2\chi \in [-\pi/2, \pi/2]$ describes the ellipticity angle (EA) of
the polarisation.\footnote{Poincar{\'e} sphere coordinates are a
  factor of two greater than the electric vector angles they
  represent. In this paper we write the factor of two explicitly so
  that all angles unambiguously refer to electric vector angles.}
Orthogonal polarisation states are represented as antipodal points on
the sphere.

Propagation effects such as Faraday rotation (FR) and generalised
Faraday rotation (GFR) are represented as rotations of the polarisation
vector on the Poincar{\'e} sphere, with trajectories tracing circles
of constant colatitude with respect to the ``modal axis'' defining the
polarisation state of the two orthogonal, natural wave modes of the
medium through which the wave is propagating \citep{Pacholczyk1970}.
In both FR and GFR the evolution occurs as a function of frequency
rather than pulse phase, and evolution along a great circle requires
an intrinsic polarisation state with a colatitude of $\SI{90}{\deg}$
with respect to the natural modes.

\subsection{Propagation through a linear birefringent medium}\label{apx:poincare-lbm}

A ``linear birefringent medium'' (LBM) is a plasma medium with two
linearly polarised natural wave modes with different phase velocities.
The natural modes can be described as a ``fast axis'' oriented at
angle $\zeta_1$ and orthogonal ``slow axis''
$\zeta_2 = \zeta_1 + \SI{90}{\deg}$, with modal axis oriented towards
longitudes of $2\zeta_1$ and $2\zeta_2$ on the equator of the
Poincar{\'e} sphere. Propagation through an LBM introduces a relative
phase delay between these two modes, characterised by a retardance
$\delta$, which depends on the plasma properties and thickness of the
birefringent medium \citeg{Lyutikov2022}. On the Poincar{\'e} sphere,
this corresponds to a rotation of the polarisation vector by an angle
$\delta$ about the modal axis.

For an initially linearly polarised wave with intrinsic PA
$\psi_{\rm in}$, the polarisation state lies on the equator of the
Poincar{\'e} sphere. If $\psi_{\rm in}$ coincides with one of the
natural modes ($\psi_{\rm in} = \zeta_1$ or $\zeta_2$), the wave
propagates in a single mode and no relative phase accumulation occurs,
leaving the polarisation state unchanged. For other intrinsic angles
$\zeta_1 < \psi_{\rm in} < \zeta_2$, the wave propagates through the
LBM in a mix of the natural modes and accumulates a relative phase,
producing a rotation away from the equator by an angle $\delta$
converting linear to elliptical polarisation.

If the intrinsic state varies as a function of pulse phase, either due
to emission geometry as with the RVM or through another mechanism, the
intrinsic polarisation vector traces an arc of a great circle along
the equator of the Poincar{\'e} sphere. Evolution along an arbitrary
great circle as a function of pulse phase can then be explained as a
rotation of the entire linearly polarised intrinsic pulse profile
about the modal axis. The observed great circle track then has an
inclination angle of $\delta$, and equatorial crossing points
occurring at longitudes corresponding to the LBM natural modes of
$2\zeta_1$ and $2\zeta_2$.

\subsection{Great circle fitting}\label{apx:poincare-gc}

We fit a great circle model to the phase-folded SB70271 pulse profile
trajectory on the Poincar{\'e} sphere using the parameterisation
\begin{equation}
\tan(2\chi) = -\tan\left(2\chi_{\max}\right)\sin\left(2\psi - 2\psi_{\rm eq}\right),
\end{equation}
where $2\chi_{\max}$ is the maximum latitude reached along the
trajectory and $2\psi_{\rm eq}$ defines the modal axis and longitude
at which the great circle intersects the equator. We used 
\textsc{emcee} \citep{Foreman-Mackey2013} to
fit this model, with best-fit posterior parameters of
$\chi_{\max} = \SI{15.8 \pm 0.3}{\deg}$ corresponding to an
inclination $\delta = 2\chi_{\max} = \SI{31.5 \pm 0.6}{\deg}$, and
$\psi_{\rm eq} = \SI{32.0 \pm 0.2}{\deg}$.

Birefringence in a magnetised plasma is typically dispersive and
wavelength dependent, with the accumulated phase difference between
the natural modes following a power law $\delta \propto
\lambda^p$. Great circles following this model are therefore expected
to show a frequency dependent inclination angle, though
$\psi_{\rm eq}$ is expected to be frequency independent and determined
purely from the modal axis of the LBM. We split the folded pulse
profile into four \SI{72}{\mega\hertz}-wide subbands and fit
independent great circles to each band. The inclination angle shows
only weak evidence for power law frequency dependence over the four
subbands, with $p_{\psi_{\rm eq,fixed}} = 0.5 \pm 0.2$ when fixing
$2\psi_{\rm eq} = \SI{64.0}{\deg}$ and
$p_{\psi_{\rm eq,free}} = 0.21 \pm 0.06$ when allowing it to vary
freely. In both cases frequency dependence of the great circle
trajectory is poorly constrained by our data, and the index implies
only mild frequency dependence for the phase delay induced by the LBM.